# Improvements on the Stability and Operation of a Magnetron H- Ion Source

A. Sosa[1], D.S. Bollinger[1], P.R. Karns[1] and C.Y. Tan[1]

[1]*Fermi National Accelerator Laboratory, PO Box 500, Batavia, IL, USA 60510*
e-mail address: asosa@fnal.gov

The magnetron H- ion sources developed in the 1970's currently in operation at Fermilab provide beam to the rest of the accelerator complex. A series of modifications to these sources have been tested in a dedicated offline test stand with the aim of improving different operational issues. The solenoid type gas valve was tested as an alternative to the piezoelectric gas valve in order to avoid its temperature dependence. A new cesium oven was designed and tested in order to avoid glass pieces that were present with the previous oven, improve thermal insulation and fine tune its temperature. A current-regulated arc modulator was developed to run the ion source at a constant arc current, providing very stable beam outputs during operations. In order to reduce beam noise, the addition of small amounts of $N_2$ gas was explored, as well as testing different cathode shapes with increasing plasma volume. This paper summarizes the studies and modifications done in the source over the last three years with the aim of improving its stability, reliability and overall performance.

## I. INTRODUCTION

C.W. Schmidt developed a version of the magnetron ion source at FNAL in the late 1970's. J. Alessi at Brookhaven National Laboratory further optimized the FNAL design, reducing the discharge current, increasing the extraction voltage and introducing a dimpled cathode that used permanent magnets [1]. Operational experience from BNL has shown that this type of source is more reliable with a longer lifetime due to better power efficiency [2]. The magnetron source design currently used at FNAL, is similar to both the BNL ion source and the source developed by C.W. Schmidt for the High Intensity Neutrino Source HINS project at FNAL [3]. The source shown in Fig. 1 has a round aperture with direct extraction using a 45° extraction cone. The source cathode has a spherical dimple which provides focusing of the H- ions leaving the surface to the anode aperture increasing the power efficiency to 48 mA/kW. The source is mounted reentrant inside a 10-inch stainless steel cube and was designed with "ease of maintenance" in mind. The operational parameters for this source are listed in Table I.

The extraction method is single stage with the extraction cone at ground potential and the source biased to -35 kV, pulsed at 15 Hz. This extraction method allows the source to run in the space charge limited regime with high extracted beam currents. The extractor pulser is a new design that uses solid state switch plates [4]. The source electronics are installed in a floating high voltage rack.

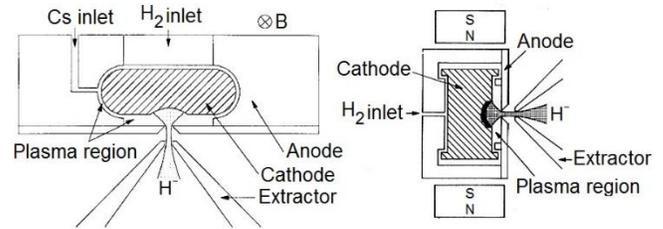

**Figure 1.** Schematic of Fermilab's magnetron H- source.

**TABLE I.** FNAL H- Source Parameters.

| Parameter | Value | Units |
|---|---|---|
| Arc Current | 15 | A |
| Arc Voltage | 180 | V |
| Extractor Voltage | 35 | kV |
| Beam Current | 80 | mA |
| Power Efficiency | 48 | mA/kW |
| Rep Rate | 15 | Hz |
| Arc Pulse Width | 250 | μs |
| Extracted Beam Pulse Width | 80 | μs |
| Duty Factor | 0.375 | % |
| Cathode Temperature | 380 | °C |
| Cs Boiler Temperature | 130 | °C |
| Emittance $\varepsilon_x/\varepsilon_y$ (norm., 95%) | 0.17/0.28 | π mm mrad |
| Extraction Gap | 4.67 | mm |
| Lifetime | 9 | months |

High frequency noise in the beam current is an issue for users of one-turn Booster operations. Significant changes in the H- population of extracted bunches propagate throughout

Linac and can be visualized using a high-bandwidth BPM [5], as shown in Fig. 2.

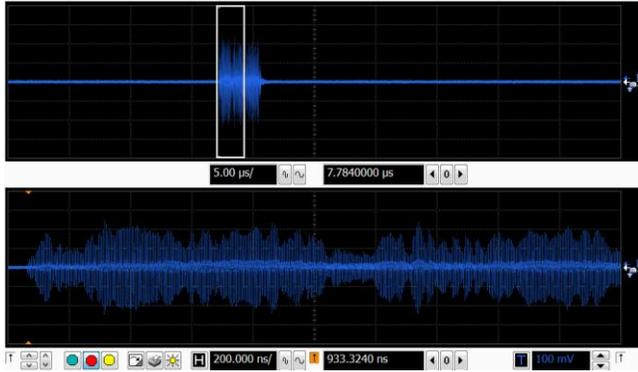

**FIG. 2.** Linac beam pulse (top) with visible 50 to 100 ns micro structures at a shorter scale (bottom).

For this reason, a set of studies were performed with the aim of reducing the beam current noise. This paper summarizes the studies and modifications done in the source over the last three years with the aim of improving its stability, reliability and overall performance.

## II. SOLENOID GAS VALVE

The typical gas pulse width used to operate the ion source is about 100 μs. An H$^-$ ion production cycle starts with injection of a single H$_2$ gas pulse into the ion source, as shown in Fig. 3, which is electrically heated to ~200°C and contains a cesiated cathode. Approximately 1 ms later, a 230 μs pulse provides -300 V for an arc discharge between anode and cathode, igniting the plasma.

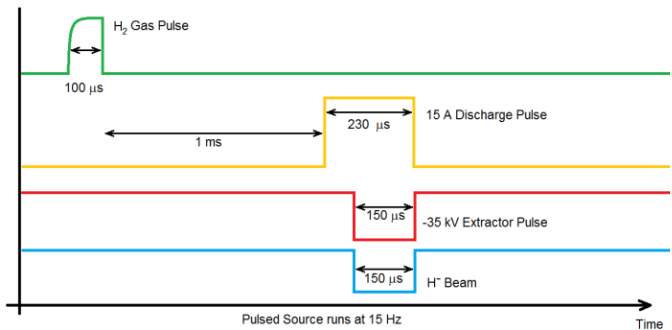

**FIG. 3.** Schematic of the H$^-$ production cycle showing the gas pulse and arc discharge.

Surface produced H$^-$ ions from the cathode are extracted with a -35 kV, 150 μs pulse synchronized with the last portion of the 230 μs arc pulse and centered on the chopped beam for the Linac. The magnetron H$^-$ ion sources currently in operation at Fermilab use Veeco PV-10 piezoelectric gas valves (Fig. 4) to pulse H$_2$ gas into the ion source [6].

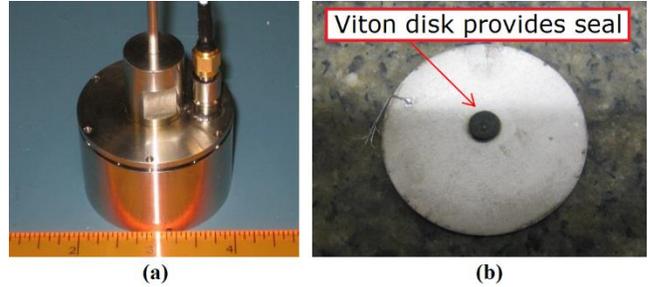

**FIG. 4.** (a) Piezoelectric valve case. (b) Piezoelectric valve shown with the Viton disk glued on it.

These valves have long been used in H$^-$ ion sources for gas injection under vacuum conditions [7, 8]. These valves open at 20 V, flexing proportionally to the applied voltage until they reach 100 V. In general these piezoelectric valves work well and are very reliable for our application but their flexibility tends to decrease with increasing temperature. As room temperature increases, the piezoelectric valve flexes less, allowing less gas into the source, thus affecting the arc current and beam output. This means a 1 °C change in room temperature can translate into a 1 μTorr change in vacuum pressure and as much as 1 A change in arc current. Operators frequently need to adjust gas valve voltage and/or gas pulse width to keep up with changes in room temperature. In addition, calibration issues with the installation of the gas valve and its tension spring account for significant performance differences from one piezoelectric valve to the next. This inconvenience motivated the need to find an alternative way of injecting H$_2$ gas into the source to avoid its temperature dependence.

A commercial pulsed solenoid valve (Series 9, Parker Hannifin Corporation, General Valve Division) has been characterized in a dedicated off-line test stand to assess the feasibility of its use in the operational ion sources. H$^-$ ion beams with beam currents >50 mA have been extracted at 35 keV using this valve. The performance of the solenoid gas valve has been characterized measuring the beam current output of the magnetron source with respect to the voltage and pulse width of the signal applied to the gas valve.

A VESPEL poppet was used due to its thermal and mechanical properties. The valve aperture is 0.5 mm in diameter. Preliminary beam runs allowed the source to operate in similar conditions as when using piezoelectric valves. The typical gas pulse width used with the piezoelectric valves is about 90 to 115 μs, compared to a range of 100 to 150 μs when using the solenoid gas valve in order to obtain similar arc current and beam output. When running the ion source with this valve, biased at 275 V and 124 μs gas pulse width, 840 μs before the arc discharge, the beam output was maximum. Vacuum pressure, extraction voltage, arc voltage and arc current remained constant during this test at 4.2 μTorr, -35 kV, 189 V and 15 A, respectively.

In these conditions, the arc pulse width was increased from 100 to 275 μs and beam current and cathode temperature were recorded. Results are plotted in Fig. 5. An arc pulse width ranging from 150 to 250 μs yielded the highest beam currents, ceteris paribus. In this range the ion source can deliver beam currents above 50 mA, which satisfies the needs of the injection line downstream.

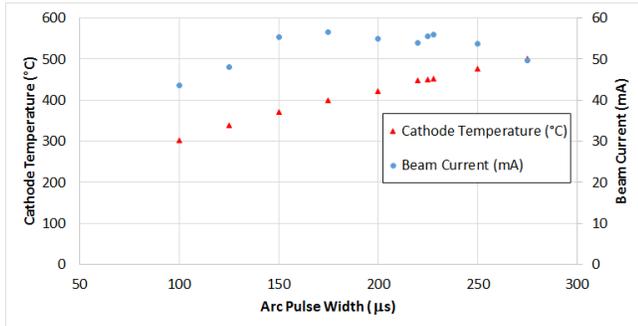

**FIG. 5.** Beam current and cathode temperature vs arc pulse width using the solenoid gas valve in the H- source.

## III. CESIUM OVEN

The magnetron ion source is a Surface Plasma Source (SPS) [9], which relies on cesium to lower the work function of the cathode for H- production [10]. A cesium oven is used to heat elemental cesium so that it flows to the rest of the cesium transport system to replenish the cesium layer on the cathode, which is depleted due to the high cathode temperature and plasma erosion. The cesium oven is used to cesiate the cathode in the ion source. The cesium ovens currently in operation at FNAL are 139.7 mm long, 14.29 mm diameter tubes made of copper (Fig. 6). A commercially available 5 g cesium vial is inserted into the copper tube, then sealed on top with an isolation valve. This assembly is then connected to the rest of the ion source.

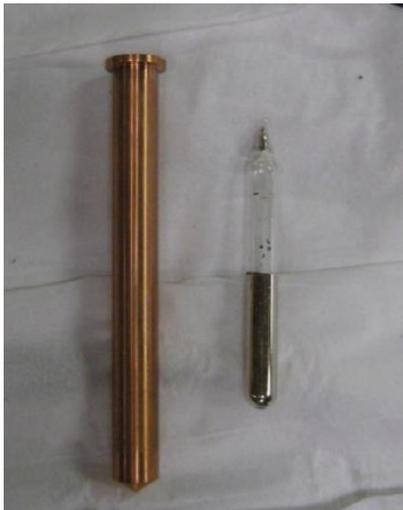

**FIG. 6.** Cesium oven made of copper (left) and vial with 5 g of elemental cesium (right).

Prior to starting an ion source with an unused boiler, the ion source is pumped down and the cesium oven is baked out with the valve open to ensure a good vacuum is achieved. Then the oven valve is closed and the copper tube is pinched with a tool in order to crack the cesium vial inside and release it. To finish, the cesium oven is wrapped up in insulating material and warmed up to operational temperatures along with the rest of the cesium transport elements going into the ion source. This method has worked well for many years, but it presents several challenges.

The cesium oven is heated using a heat tape made by BriskHeat® Corporation [11]. The excess heat tape was wrapped around both the valve and cesium transport tube. This meant that the temperature of the boiler, valve and tube were all interactive so it was hard to balance the difference in temperatures. Another issue with the cesium oven is that, copper being a very good heat conductor made it hard to obtain the low nominal oven temperature of 110°C as the cesium oven is heated not only by the heat tape, but by the valve heater as well. The heat tape had to be run at very small currents along with thinning out the exterior cesium oven insulation. As a result, the cesium oven temperature would start following the room temperature.

A new cesium oven made out of stainless steel was designed to fix the aforementioned problems (Fig. 7). It is 63.5 mm long and 19 mm in diameter. The more compact design helps with thermal insulation of the cesium oven from the rest of the cesium delivery systems. The smaller design also requires pouring the cesium into it from the glass vials by means of a glove bag purged with $N_2$. Stainless steel was chosen as a material for the new oven because, although stainless steel has about 30 times less thermal conductivity, it avoids heat dissipation from the oven into the sealing valve sitting on top of it.

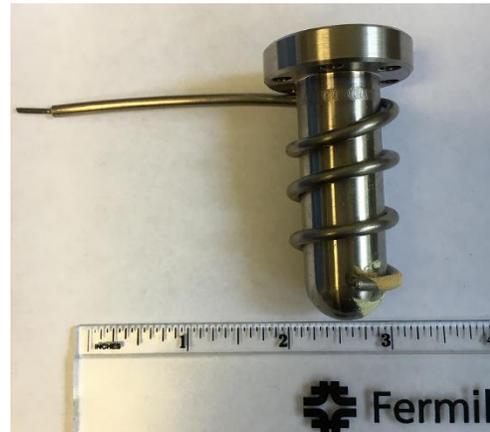

**FIG. 7.** New cesium oven made of stainless steel with 2 Ω/m heating wire around it.

In order to select a suitable heating wire for the new oven, a quick test was done comparing the copper and stainless-steel ovens with two different types of wire: 5.33 Ω/m and 2 Ω/m. These wires are AerOcoax® made by ARi Industries Inc [12]. In Fig. 8 the temperature response of the copper oven is shown with 4 turns of 5.33 Ω/m wire and 10 turns of 2 Ω/m wire. The line for the stainless steel oven corresponds to 4 turns of 2 Ω/m wire, as the 5.33 Ω/m wire takes very little current to heat up resulting in a slope too steep to be useful.

In our application it is desirable to have the ability to fine tune the temperature (ideally to less than 1°C) for every step in current from the power supply. Although the copper oven with 4 turns of the 2 Ω/m wire has the smallest slope of the three, the stainless steel oven also provides the ability to fine tune the temperature with small current increments as well as the benefits of avoiding glass fragments from the glass vial and better thermal insulation.

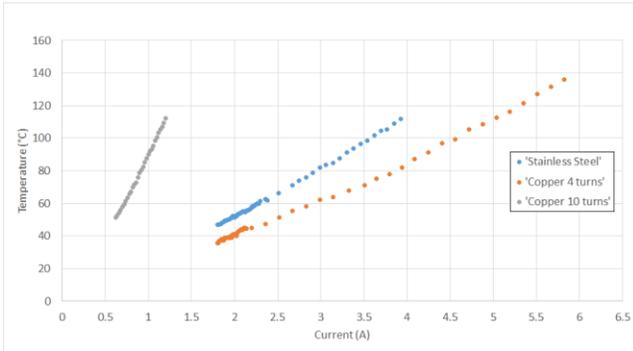

**FIG. 8.** Temperature response of the different oven configurations as a function of current.

## IV. CURRENT-REGULATED ARC MODULATOR

The arc pulser currently used in the operational sources is voltage regulated. The main inconvenience of running an ion source with a voltage-regulated arc modulator is the drift of the arc current over time. Since arc current and beam output are intimately related, a drifting arc current means an unstable beam output over the course of beam operations. In order to avoid this, a Current-Regulated (CR) arc modulator was designed at FNAL, inspired by the CR arc modulator from BNL [13]. It consists of a high-voltage card and a trigger/feedback loop card as shown in Fig. 9.

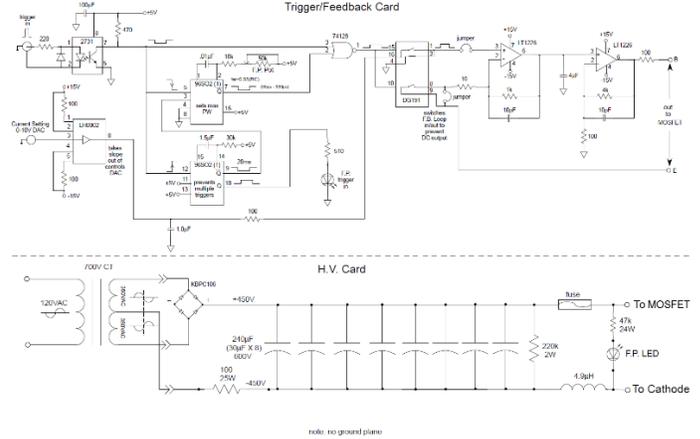

**FIG. 9.** Circuit diagram of the CR arc modulator.

The voltage to current converter comprises an op-amp set up as a positive high gain amplifier connected to a power MOSFET which acts as a gate for controlling the large current flowing out of the high voltage power supply.

The high current power MOSFET IXFN132N50P3 [14] is spark resistant at -35 kV. It has a forward DC current rating of 112 A and a drain-source voltage rating of 500 V. The fast rise and fall times of ~100 ns of this MOSFET help significantly reduce the fall and rise times of the arc current pulse. Notice that the high-voltage card is floating (not connected to the ground reference) as shown in Fig. 9. The different components of the CR arc modulator and their layout in the chassis are shown in Fig. 10.

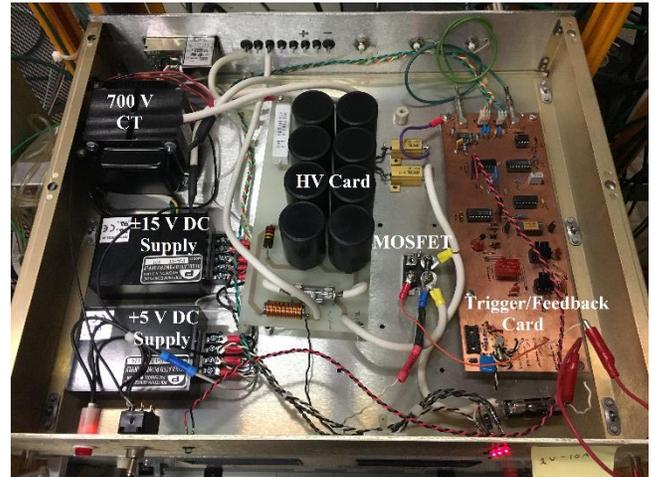

**FIG. 10.** CR arc modulator chassis showing the 700 V transformer, DC power supplies, high-voltage card, trigger/feedback card and the MOSFET.

In Fig. 11 the performance of the CR-modulator over 3 days of beam time is shown. The arc current is very steady at 15 A even though the arc voltage and vacuum pressure change over time. This represents a major improvement towards more steady beam outputs during operations.

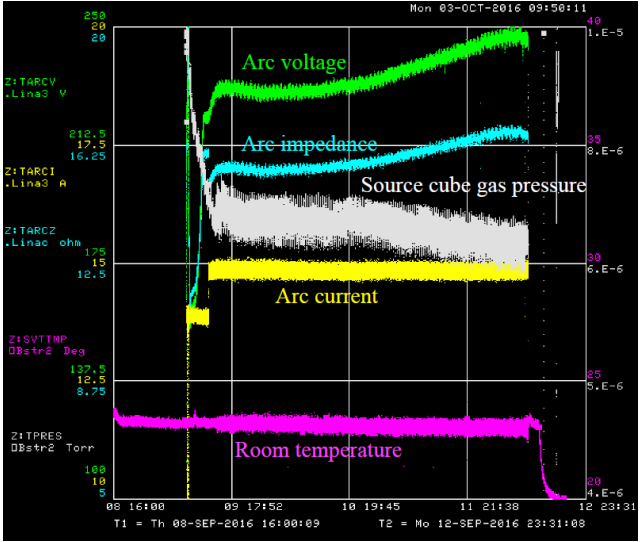

**FIG. 11.** Performance of the CR-modulator over 3 days running in the ion source. The arc current remains constant whereas the arc voltage and vacuum pressure wander.

Further studies will be done when the system is installed in an operational system since the effects of changing source impedance on downstream machines is unknown yet. Table II summarizes the design specifications for the CR-modulator.

**TABLE II. Arc Modulator Specifications**

| Parameter | Value | Units |
|---|---|---|
| Pulse Width | 100-300 | μs |
| Rise/Fall Time | < 20 | μs |
| Regulated Current Range | 1-20 | A |
| Current Ripple Amplitude | < 1 | % |

## V. BEAM CURRENT NOISE

Earlier studies have been performed exploring different methods of reducing the beam current noise of the magnetron [15]. It has been previously suggested that adding a small amount (1% in volume) of $N_2$ gas decreases the modulation or noise of the extracted beam current [16, 17]. An experiment was done in the offline test stand to mix $H_2$ and $N_2$ gases very accurately at different ratios on the gas line that feeds the source with the aim of exploring the effect of different $N_2/H_2$ ratios on the beam current noise.

In this study, beam current noise was defined as the ratio (as a percentage) between the standard deviation and the mean signal from the beamline toroid:

$$Noise\ Ratio = \frac{\sigma}{\bar{I}} \times 100 \quad (1)$$

The experimental setup is depicted in Fig. 12. Two Mass Flow Controllers (MFCs) which are calibrated for each pure gas regulate the amount of each gas flowing into the source. These MFCs measure pressure drop in laminar flows of gas (10 to 15 psi) which make them accurate down to 0.8% of the reading and 0.2% of their full scale [18]. The full scale of the $N_2$ MFC is 0 to 0.5 sccm whereas that of the $H_2$ MFC is 0 to 5 sccm. With this setup, $N_2/H_2$ ratios between 0.5 to 3% could be tested, and the mean beam current noise was measured connecting the output of the current toroid to an oscilloscope.

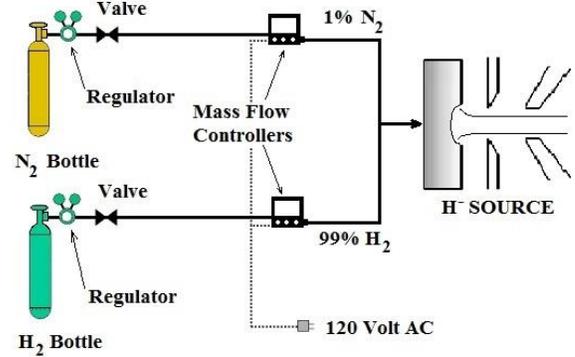

**FIG. 12.** Experimental setup for gas mixing into the ion source.

Fig. 13 shows a scope picture of the high frequency noise of the beam current signal from the downstream toroid. Noise ratio was calculated from the values measured within a 50 μs window in the signal flat top.

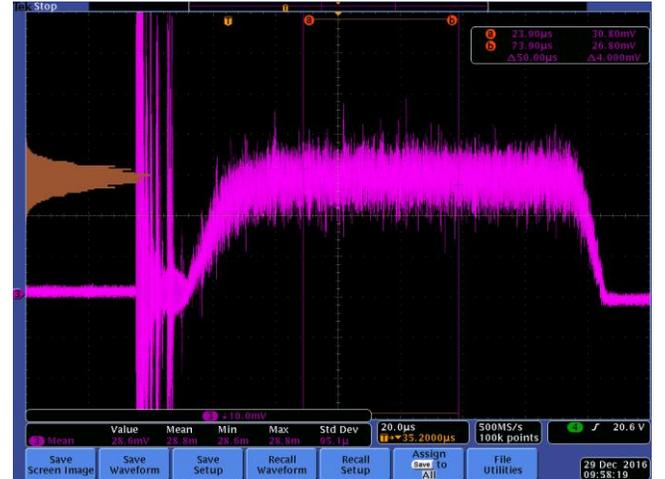

**FIG. 13.** Scope picture of the beam current signal from the toroid downstream of the source. Beam extracted at -35 kV.

In Fig. 14 the noise ratio of a magnetron ion source with a regular dimpled cathode is shown as a function of the percentage of nitrogen. For every data point, the maximum beam current was recorded at an extraction voltage of -35 kV. A slight increase in noise ratio is seen as more nitrogen is added to the source, however the noise ratio is very low in every data point, at under 1%. The observed trend in the data also reflects that, as more nitrogen is added to the source, the beam current signal decreases while the standard deviation of the signal remains unchanged, hence the increase in the noise ratio. No significant reduction in the noise ratio was observed in these conditions. Above 3% $N_2$ the arc current is severely affected or even terminated in a matter of minutes.

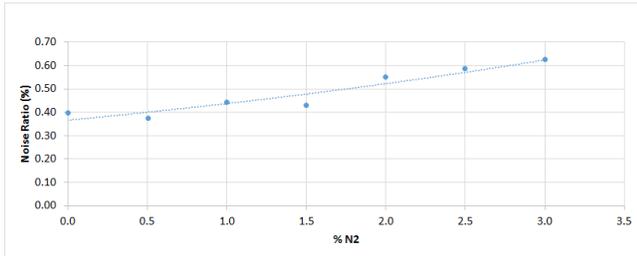

**FIG. 14.** Noise ratio as a function of $N_2/H_2$ ratio.

In Fig. 15 the noise ratio was studied for different arc currents at the same extraction voltage of -35 kV. In general, the extracted beam current increases linearly with arc current, therefore a lower noise ratio is observed at higher arc currents with and without $N_2$. No significant difference is observed in the plot by adding $N_2$ to the source. From the data set, the only magnitude that changes appreciably is the beam current signal but not its standard deviation.

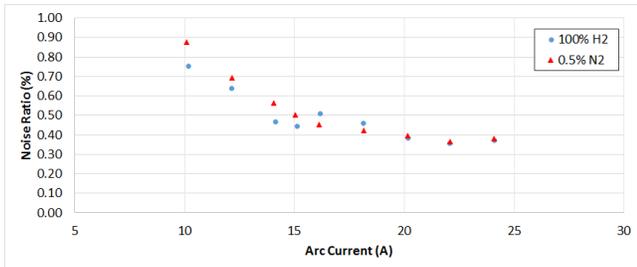

**FIG. 15.** Noise ratio as a function of arc current with and without $N_2$ in the gas mixture.

The effect of beam current noise was also studied for different types of cathodes. The main hypothesis is to verify whether a smaller source cathode, which allows for a larger plasma volume in a magnetron source yields a smaller noise ratio.

The cathodes used in this test were: the standard dimple cathode, the hollow cathode (see description in next section), a thin cathode and a cathode with the dimple drilled through. These cathodes are illustrated in Fig. 16.

The standard cathode (Fig. 16a) has a dimple 8.4 mm in diameter and 1.59 mm deep. This cathode is 9.04 mm thick from front to back. The so-called hollow cathode is just a standard cathode with a hole 2.9 mm in diameter and 5.54 mm deep on one side. The thin cathode (Fig. 16c) is 6 mm thick and has a groove 0.3 mm deep on the front side. The drilled through cathode is a standard cathode with a hole 8.41 mm in diameter drilled all the way through the cathode.

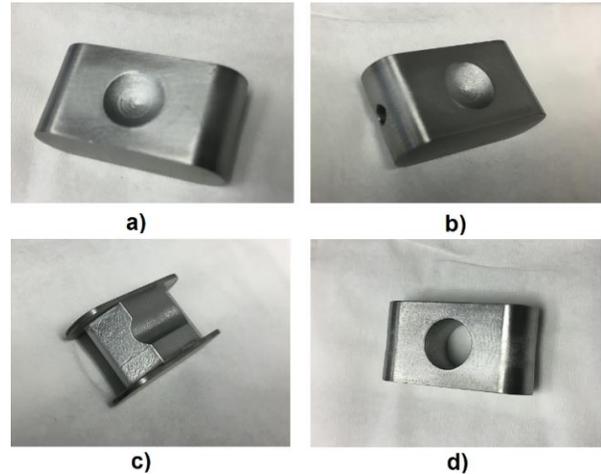

**FIG. 16.** Cathodes used in the ion source: a) Standard dimpled cathode b) Hollow cathode c) Thin cathode d) Drilled through cathode.

The ion source was installed with each cathode described above and run to a stable operational point with the extractor voltage set at the nominal value of -35 kV. With the CR arc modulator, the arc current was set anywhere from 10 to 22 A and the noise ratio was recorded by measuring the mean current signal from the toroid and its standard deviation. When plotting the noise ratio as a function of arc current (Fig. 17), no significant trend can be observed from the data except for the thin cathode, and noise ratios remain relatively low (<2%) for all cathodes in this arc current range. The hollow cathode performs best for arc currents >15 A.

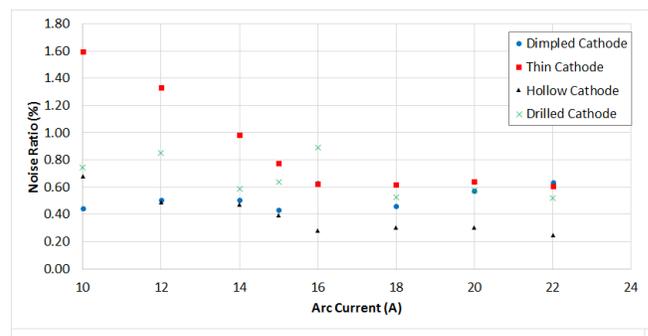

**FIG. 17.** Noise ratio as a function of arc current for different cathodes. Extraction voltage is -35 kV.

In addition, the average pressure in the source chamber was changed from 4.5 to 10 µTorr by adjusting the gas valve voltage to allow more or less $H_2$ into the source. In Fig. 18, noise ratio data for all the different cathodes was recorded while running the source at 15 A of arc current and extracting at -35 kV. A few data points could not be taken with certain cathodes at high pressure due to excessive sparking.

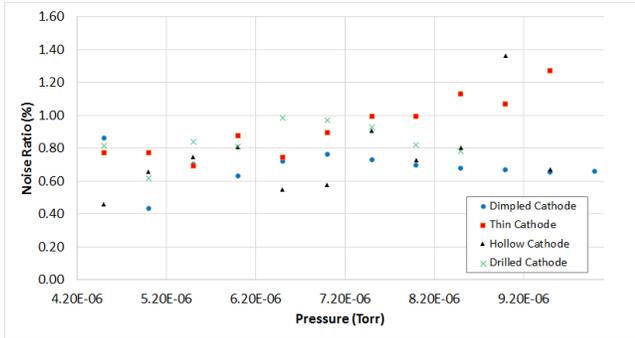

**FIG. 18.** Noise ratio as a function of pressure for different cathodes. Arc current is 15 A and extraction voltage is -35 kV.

From this data no clear pattern indicates a change in noise ratio for different cathodes within this pressure range. Note that the noise ratio remains low (<2%) for all cathodes in this pressure range.

In the case of the drilled through cathode not many surface produced ions can be extracted, as no surface is directly facing the exit aperture of the source. In this condition, the argument can be made that most of the extracted ions come from the plasma volume between cathode and anode. If that is the case one can compare the low beam currents (<20 mA) extracted from the drilled cathode to the rest of cathodes. This would imply that surface produced ions account for 80% of the extracted beam current.

## VI. CONCLUSIONS

When using the solenoid gas valve and an arc pulse width of 150 to 250 µs, the magnetron ion source can deliver beam currents above 50 mA, which satisfies the needs of the injection line downstream. This solenoid gas valve was observed to be spark resistant in addition to running stably with no observed correlation to changes in room temperature.

The new stainless steel cesium oven provides the ability to fine tune the temperature with little current increments. In addition, it avoids the glass fragments from the broken ampoule and provides easier thermal insulation with a reduced oven surface. This means that room temperature has less effect on the boiler temperature.

The CR arc modulator represents a major improvement towards more steady beam outputs during operations. Very steady arc currents can be achieved over days of beam run with minor offsets (<1 A) relative to the set point defined by the user. During operation in current-regulated mode, the arc voltage and arc impedance can swing accordingly while providing a stable beam current output. Further studies will be done in an operational system to evaluate the effects of changing source impedance on downstream machines.

None of the proposed ideas resulted in reduced beam current or arc discharge noise. The magnetron ion source is known to be inherently noisy. As shown in Fig. 19, the gas density n required for the magnetron ion source in order to achieve noiseless discharge is above the gas density $n_m$ in which extraction voltage breakdown and/or ion stripping is high [19]. For noiseless discharge, the magnetic field in the magnetron source would need to be less than 1 kG, but in that case the extracted beam current drops and the amount of co-extracted electrons increases sparking. This renders the noiseless discharge regime of the magnetron in an unprofitable region to deliver beam operationally.

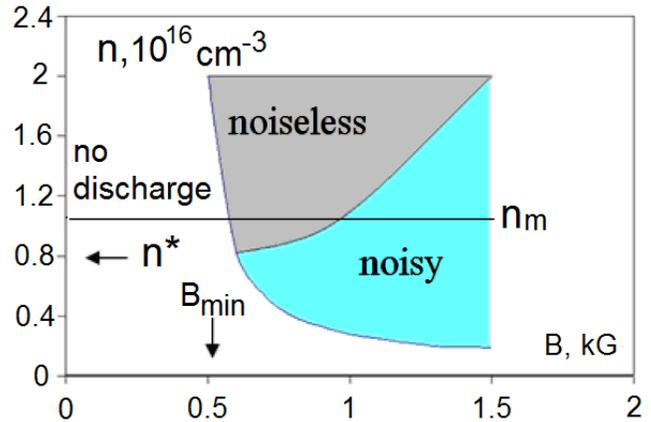

**FIG. 19.** Diagram of the magnetron discharge stability as a function of magnetic field B and gas density n.

A viable alternative would be installing a Penning source, which is known to achieve noiseless arc discharge [19], which would translate into a reduction of the extracted beam current noise.

The Penning is the brightest H- ion source with current densities at extraction above 1 A cm$^{-2}$. The lifetime of the Penning source is limited to a few weeks because of cathode sputtering by cesium ions. This type of source will not operate without cesium vapour and requires the electrode surfaces to be between 400°C and 600°C. This source has been developed by D. Faircloth and his team at Rutherford Appleton Laboratory. 60 mA, 1 ms, 50 Hz pulses can be routinely produced with this source [20].

### 1. ACKNOWLEDGEMENTS

This research was supported by Fermi Research Alliance, LLC under Contract No. De-AC02-07CH11359 with the United States Department of Energy. I would like to thank A. Feld and K. Koch for their technical support with the gas valves and the ion source during these experiments. I also want to thank T. Lehn (BNL) for sharing the design notes of their current regulated arc modulator.